\documentclass[conference]{IEEEtran}
\IEEEoverridecommandlockouts
\usepackage[binary-units]{siunitx}
\usepackage{tabularx}
\usepackage{multirow}
\usepackage{color, colortbl}
\usepackage[binary-units]{siunitx} 
\DeclareSIUnit{\bps}{bps}
\usepackage{graphicx}
\usepackage{cite}
\usepackage{amssymb}
\usepackage{amsmath}
\usepackage[caption=false,font=footnotesize]{subfig}
\def\BibTeX{{\rm B\kern-.05em{\sc i\kern-.025em b}\kern-.08em
    T\kern-.1667em\lower.7ex\hbox{E}\kern-.125emX}}
\usepackage{flushend}

\begin{document}
\title{Terahertz-Enpowered Communications and Sensing in 6G Systems: Opportunities and Challenges }

\author{\IEEEauthorblockN{Wei Jiang\IEEEauthorrefmark{1} and Hans D. Schotten\IEEEauthorrefmark{2}}
\IEEEauthorblockA{\IEEEauthorrefmark{1}German Research Center for Artificial Intelligence (DFKI)\\Trippstadter Street 122,  Kaiserslautern, 67663 Germany\\
  }
\IEEEauthorblockA{\IEEEauthorrefmark{2}Rheinland-Pf\"alzische Technische Universit\"at Kaiserslautern-Landau\\Building 11, Paul-Ehrlich Street, Kaiserslautern, 67663 Germany\\
 }
}

\maketitle
\begin{abstract}
The current focus of academia and the telecommunications industry has been shifted to the development of the six-generation (6G) cellular technology, also formally referred to as IMT-2030. Unprecedented applications that 6G aims to accommodate demand extreme communications performance and, in addition, disruptive capabilities such as network sensing. Recently, there has been a surge of interest in terahertz (THz) frequencies as it offers not only massive spectral resources for communication but also distinct advantages in sensing, positioning, and imaging. The aim of this paper is to provide a brief outlook on opportunities opened by this under-exploited band and challenges that must be addressed to materialize the potential of THz-based communications and sensing in 6G systems. 
\end{abstract}

\IEEEpeerreviewmaketitle

\section{Introduction}

\begin{figure}[!bpht]
\centering
\includegraphics[width=0.3\textwidth]{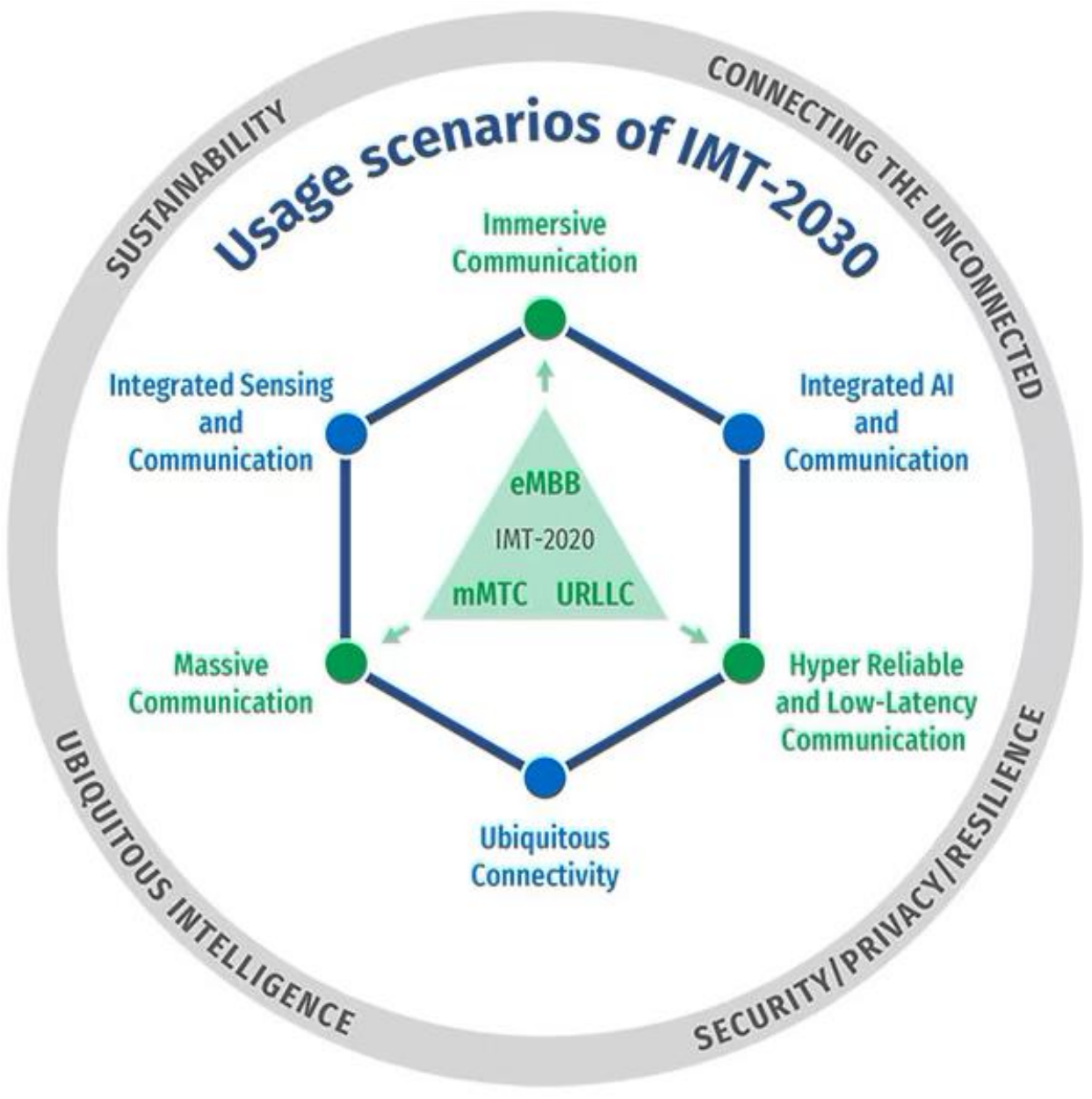}
\caption{Six usage scenarios for IMT-2030 and four overarching aspects.}
\label{Figure_6GScenarios}
\end{figure}

Starting from 1985, when IMT-2000, also known as 3G, was developed, the ITU-R has been actively engaged in the standardization of each generation of cellular systems. Like the definition of visions for 4G (IMT-Advanced) and 5G (IMT-2020), in the form of recommendations ITU-R M.1645 \cite{Ref_itu20034Gvision} and ITU-R M.2083 \cite{Ref_WJ_non2015imt}, respectively, the ITU-R started the development process for 6G by defining the vision of IMT-2030 as the first step \cite{Ref_jiang2021kickoff}. In early 2021, significant progress was made with regard to this initiative when ITU-R WP 5D formally started the study for the new recommendation referred to as ITU-R M.[IMT Framework for 2030 and Beyond]. After over two years of discussions, the draft of the new recommendation for the IMT-2030 vision was completed on 22nd June 2023 during the 44th ITU-R WP 5D meeting in Geneva. This Framework recommendation specifies six usage scenarios of IMT-2030, as shown in \figurename \ref{Figure_6GScenarios}, including \textit{Immersive Communication}, \textit{Hyper Reliable and Low-Latency Communication}, \textit{Massive Communication}, \textit{Ubiquitous Connectivity}, \textit{Integrated AI and Communication}, and \textit{Integrated Sensing and Communication}.

Driven by technological advances, 6G is expected to integrate sensing and communications into a unified system \cite{Ref_jiang2021road}. The usage scenario of \textit{Integrated Sensing and Communication} facilitates new applications and services that require sensing capabilities, which makes use of IMT-2030 to offer wide area multi-dimensional sensing that provides spatial information about unconnected objects as well as connected devices and their movements and surroundings. 
The terahertz (THz) spectrum has garnered significant attention in recent years and is being identified as a promising avenue for 6G development \cite{Ref_rappaport2019wireless}. The massive spectrum available within the THz frequencies presents prospects for high-speed wireless applications. Due to the tiny wavelengths of THz signals, antennas can be miniaturized, thereby unlocking possibilities for innovative applications like nanoscale communications \cite{Ref_lemic2021survey}. Moreover, THz signals extend their utility beyond communication by facilitating high-definition sensing, imaging, and precise positioning within the immediate physical surroundings. 

The objective of this paper is to provide a brief outlook on opportunities opened by this under-exploited band and challenges that must be addressed to materialize the potential of THz-based communications and sensing. The next section will elaborate on the potential uses of THz frequencies in 6G communications and sensing, while the technical issues will be listed in Section III.

\section{THz Opportunities in 6G}
This paper would like to clarify a fundamental concern that might still cause confusion or disputes in this field. That is, \textit{What is the rationale behind exploiting the THz band in 6G?} We address this concern through shedding light on the opportunities opened in THz communications, THz sensing, and the integration between THz communications and THz sensing.

\subsection{Immersive Tbps Communication}
As one of the 6G usage scenarios, immersive communication extends the enhanced mobile broadband (eMBB) of 5G, with a primary focus on delivering immersive experiences for users. It supports a wide range of use cases and applications such as XR, remote multi-sensory telepresence, and holographic services. The realization of these captivating experiences might necessitate exceptionally high data transfer speeds, potentially on the order of magnitude of Tbps. 

As we know, the electromagnetic spectrum consists of radio, microwave, infrared, visible light, ultraviolet, X-rays, and Gamma rays, from the lower to higher frequencies. THz is considered a suitable candidate to realize Tbps communications under the current level of hardware and signal-processing technologies. The reasons are explained as follows: 
\begin{itemize}
    \item \textit{Spectrum scarcity of the sub-\SI{6}{\giga\hertz} band}: Favourable propagation characteristics of sub-\SI{6}{\giga\hertz} frequencies facilitate the use of sophisticated transmission technologies such as massive MIMO \cite{Ref_jiang2021cellfree}, non-orthogonal multiple access, and high-order modulation like 1024QAM to achieve high spectral efficiency. However, spectrum scarcity and non-continuity pose a significant challenge to achieving higher rates.   
    \item \textit{Implementation difficulties in mmWave bands}: By far,  there are a total of \SI{13.5}{\giga\hertz} spectrum below \SI{100}{\giga\hertz}, which was assigned by the ITU-R at the World Radiocommunication Conference 2019 (WRC-19) for the deployment of mmWave 5G systems. A data rate of \SI{1}{\tera\bps} can only be achieved with transmission schemes having a spectral efficiency of approaching \SI{100}{\bps/\hertz}, which is currently challenging to implement for mmWave signals or can be reached but limited in special conditions. 
    \item \textit{Obstacles of the optical bands}: Despite the massive spectrum available in optical bands encompassing infrared, visible-light \cite{Ref_pathak2015visible}, and ultraviolet frequencies, several challenges curtail their practical utilization. Constrained transmission power due to hardware limitations and safety concerns, the adverse effects of various atmospheric attenuations on signal propagation (e.g., fog, rain, dust, or pollution), significant losses caused by diffuse reflection, and the detrimental impact of misalignment between transmitters and receivers collectively impose restrictions on data rates and transmission distances. 
    \item \textit{Adverse effects of extreme high bands}: 
    Ionizing radiation, encompassing ultraviolet, X-rays, and Gamma rays, presents a noteworthy hazard to human well-being due to its potent energy capable of displacing electrons and initiating the formation of free radicals, which in turn can result in cancer. While the detrimental health repercussions of ionizing radiation can be managed through cautious application, its risk level remains prohibitive for personal communication purposes \cite{Ref_rappaport2019wireless}.
\end{itemize}

THz frequencies remain non-ionizing due to their lower photon energy, substantially weaker than the energy levels required for ionization in ionizing radiation. The THz band presents an extensive spectrum, spanning from tens of gigahertz to several terahertz, contingent upon the transmission distance. This grants a bandwidth exceeding that of mmWave bands by over tenfold, while the operating frequency remains at least one order of magnitude below that of optical bands. Moreover, the technological advancements essential for achieving \si{\tera\bps} transmission across the THz band are progressing rapidly. 

\subsection{THz Nano-Networks}

As commonly understood, the minimal dimensions of an antenna suitable for transmitting THz signals can fall within the range of micrometers. Conceptually, this opens the door to wireless connectivity among nanoscale machines or nanomachines, facilitated by the deployment of nanoscale antennas. A number of distinct scenarios for THz nano-communications are as follows 
\begin{itemize}
    \item \textit{Health Monitoring:} Nanoscale biosensors introduced into the human body or positioned beneath the skin have the capability to identify substances like sodium, glucose, various ions in the blood, cholesterol levels, cancer biomarkers, and the existence of diverse infectious agents. These miniature biosensors, when distributed within the body or positioned in proximity, form a body sensor network. This network has the potential to gather pertinent physiological and biochemical information concerning an individual's health status. 
    \item \textit{Nuclear, Biological, and Chemical Defense:} Distributed across their surroundings, chemical and biological nanosensors exhibit the capability to identify noxious chemicals and biological hazards. A primary advantage of employing nanosensors, as opposed to traditional macroscale or microscale sensors, lies in their ability to detect even minute concentrations of a chemical compound, reaching as low as a single molecule. Furthermore, nanosensors facilitate considerably faster detection compared to conventional sensor systems. 
    \item \textit{Internet-of-Nano-Things:} Leveraging THz nano-communications to establish connections between nanoscale machines, devices, and sensors and integrating them with current wireless networks and the Internet results in a bona fide cyber-physical system known as the Internet of Nano-Things (IoNT). This transformative framework paves the way for groundbreaking applications that have the potential to fundamentally alter human work and lifestyle paradigms.
    \item \textit{On-Chip Communication:} Utilizing THz communications presents a viable and scalable strategy for establishing inter-core connections within on-chip wireless networks. This is achieved through the implementation of planar nano-antenna arrays, which enable the establishment of ultra-high-speed links. 
\end{itemize}

\subsection{THz Sensing, Imaging, and Positioning}
As frequencies rise, the propagated signal's spatial resolution becomes increasingly refined, consequently facilitating a heightened level of precision in spatial differentiation \cite{Ref_sarieddeen2020next}. In comparison to wireless sensing across alternative frequency bands, THz sensing provides the subsequent advantages:
\begin{itemize}
    \item \textit{High resolution and penetration capability}: While low-frequency signals, such as those utilized in radar and GNSS, have the capacity to sense, detect, and pinpoint objects, THz sensing/positioning presents an opportunity for enhanced resolution, thanks to its tiny wavelengths. This advantage extends even to objects concealed from direct line of sight. THz waves possess the ability to permeate various non-conductive materials, including plastics, textiles, paper, ceramics, and dielectric substances. 
    \item \textit{Non-ionizing radiation}: In contrast to X-rays and Gamma rays, THz waves possess significantly reduced photon energy, rendering them non-ionizing. Consequently, THz sensing is widely acknowledged as safe for biological specimens and human subjects. This characteristic permits the utilization of THz waves for imaging and diagnostics that are both non-destructive and non-intrusive in nature.
    \item \textit{Low environmental interference}: Unlike visible or infrared radiation, THz waves are less susceptible to environmental factors such as ambient light, fog, or smoke. This resilience enables THz sensing to operate effectively in outdoor or challenging conditions, thereby broadening its applicability in areas such as remote sensing, atmospheric monitoring, and outdoor security.
    \item \textit{Spectroscopic analysis}: THz waves exhibit distinct interactions with molecules, resulting in the creation of characteristic spectral fingerprints. The field of THz spectroscopy offers essential insights into molecular vibrations and rotational transitions, thereby facilitating the discernment and examination of various chemical compounds, including explosives, pharmaceuticals, and biomolecules. 
\end{itemize}

\subsection{Integrated THz Communications and Sensing} 
As previously discussed, the THz band not only offers massive spectral resources for wireless communications but also presents distinct advantages in sensing, positioning, imaging, and spectroscopy \cite{Ref_sarieddeen2020next}. This has sparked considerable interest recently, positioning it as a pivotal facilitator for the implementation of Integrated Sensing and Communications (ISAC) in the context of 6G and beyond. By amalgamating THz communications and THz sensing into a cohesive framework, these dual-functional wireless networks foster a synergistic relationship through the concepts of "sensing-aided communication" and "communication-aided sensing." Leveraging sensing information within the communication process stands out as a noteworthy advantage of ISAC. It brings forth a more predictable and deterministic propagation channel, thereby facilitating the development of efficient communication algorithms and protocols. These include approaches like sensing-aided channel estimation, predictive beamforming supported by sensing data, rapid beam alignment and tracking, as well as strategies for mitigating link blockage \cite{Ref_wang2018internet}. Conversely, conventional mobile communication networks also offer significant avenues for network-based sensing or sensing as a service. Nodes within the network collaborate to share sensing results, wherein multiple network components such as base stations and user equipment collectively function as an integrated sensing system. This collaborative effort, achieved through the fusion of sensing data, serves to diminish measurement uncertainty, enhance coverage area, and elevate sensing precision and resolution.

\section{Major Challenges}

While the considerable promise of THz communications and sensing in the context of 6G is acknowledged, there exists a series of technical hurdles that must be tackled prior to the effective deployment of THz technology. This section will elucidate the primary challenges that remain unresolved and require dedicated research endeavors.

\subsection{High Free-Space Path Loss}
When an isotropic radiator projects an electromagnetic wave into open space, the energy spreads uniformly across the expanding surface of a sphere. The parameter known as \textit{Effective Isotropic Radiated Power (EIRP)} signifies the maximum energy emitted in a specific direction compared to a theoretical isotropic antenna with a unit gain. Consequently, it is the result of multiplying the transmission power by the transmitting antenna's gain in the direction of the receiving antenna.  The density of power flux, representing the power per unit area of the incident field at the antenna, is equivalent to the EIRP divided by the surface area of a sphere with that radius. As a result, the Free-Space Path Loss (FSPL) substantially grows with the increase of frequencies \cite{Ref_jiang20226GCH5}. 

\subsection{Gaseous Absorption}

Though gaseous molecules do absorb a portion of electromagnetic wave energy, the impact of atmospheric absorption on frequencies below 6GHz is so negligible that conventional cellular systems do not factor it significantly when computing the link budget. However, this phenomenon becomes remarkably pronounced in the case of THz waves, where absorption losses escalate dramatically at specific frequencies. This attenuation phenomenon emerges from the interaction between an electromagnetic wave and a gaseous molecule. When the wavelength of THz waves approaches the size of atmospheric molecules, the incoming wave induces rotational and vibrational transitions within polar molecules. These processes possess quantum characteristics, featuring resonant behaviors at distinct frequencies based on the inherent molecular structure. This leads to substantial absorption peaks at specific frequencies \cite{Ref_slocum2013atmospheric}.

\begin{figure}[!tbph]
\centering
\includegraphics[width=0.48\textwidth]{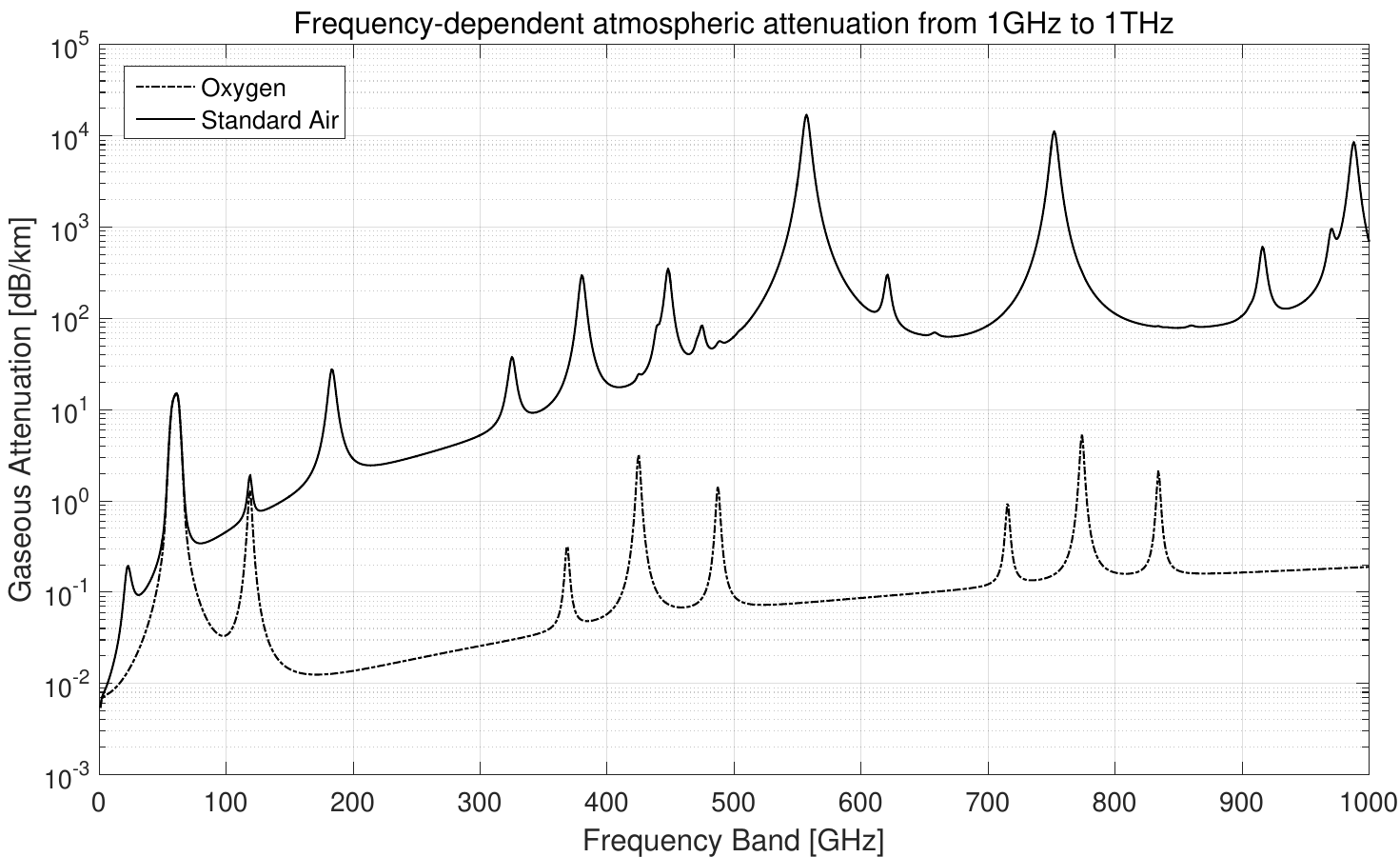}
\caption{Illustration of atmospheric absorption, according to \cite{Ref_WJ_itu2019attenuation}. } 
\label{Diagram_atmosphericAbsorption}
\end{figure}

Being a principal constituent of the atmosphere, oxygen assumes a critical role in atmospheric absorption, particularly in conditions of clear air. Simultaneously, the presence of water vapor suspended in the air exerts a potent influence on the propagation of electromagnetic waves. The damping effects attributable to water vapor take precedence in the THz band, except within a few specific spectral zones where the impact of oxygen becomes more conspicuous. While a more comprehensive exploration of atmospheric absorption is commonly conducted in fields like radio astronomy and remote sensing, the context of wireless communications generally considers the absorption effects of certain additional molecular species—such as oxygen isotopic species, vibrationally excited oxygen species, stratospheric ozone, isotopic ozone species, vibrationally excited ozone species, various nitrogen, carbon, and sulfur oxides—as minimal in comparison to the absorption caused by water vapor and oxygen. The atmospheric attenuation from \SIrange{1}{1000}{\giga\hertz} is illustrated in Fig.~\ref{Diagram_atmosphericAbsorption}.

\subsection{Weather Effects}

Apart from gaseous absorption, another atmospheric factor to consider in outdoor settings is the influence of weather.  Similar to the impact of water vapor, these studies have highlighted that suspended liquid water droplets, present in forms such as clouds, fog, snowflakes, or falling rain, can absorb or scatter incident signals. This is due to their physical dimensions aligning with the scale of THz wavelengths. While this attenuation isn't as prominent as path loss or atmospheric absorption, it remains a crucial consideration for accurate channel characterization \cite{Ref_weng2021millimeter}.

\begin{figure}
\centering
\includegraphics[width=0.47\textwidth]{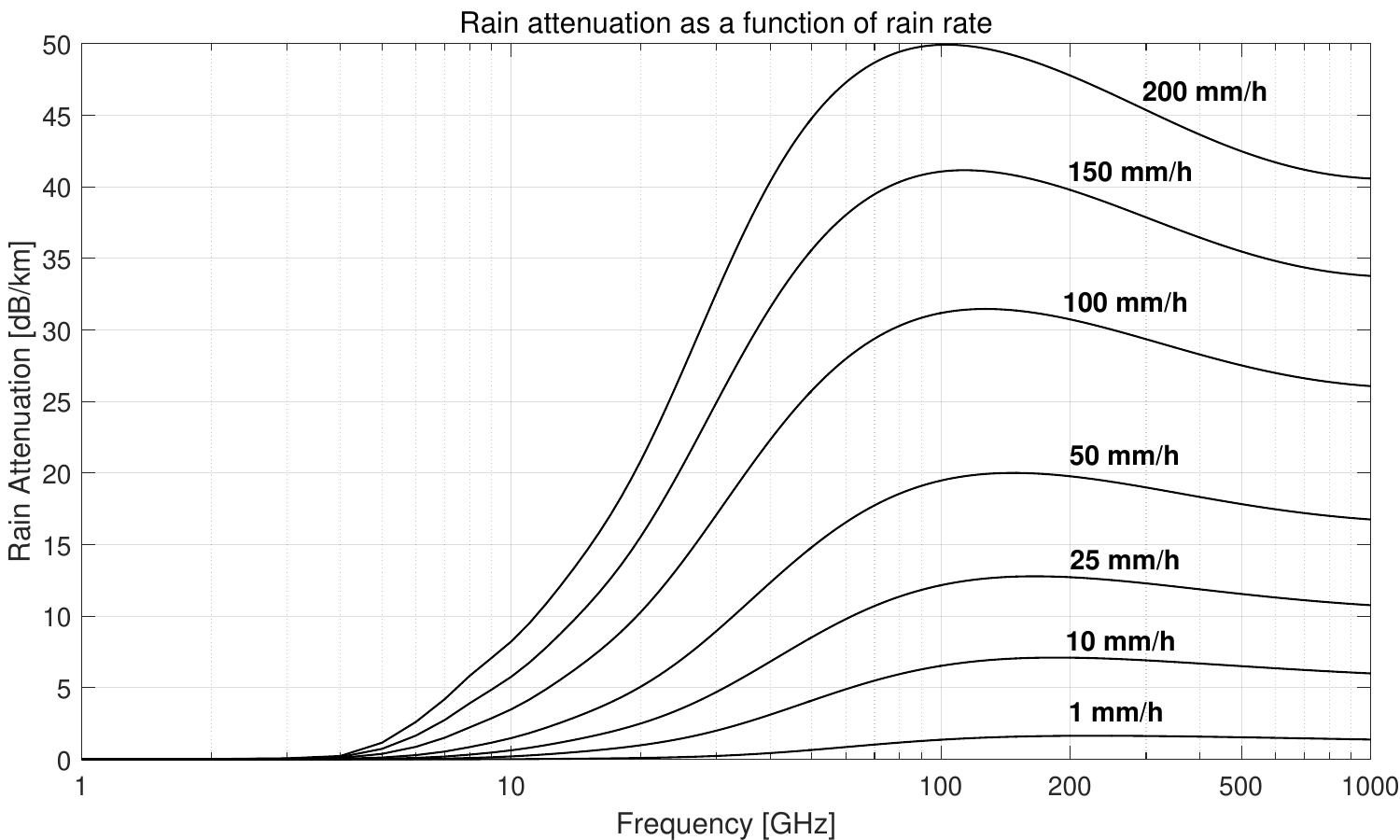}
\caption{Rain attenuation for terrestrial communications links as a function of rain rate and frequency.} 
\label{Diagram_RainAttenu}
\end{figure}

Clouds consist of minute water particles, some as tiny as \SI{1}{\micro\meter}, or ice crystals ranging from \SIrange{0.1}{1}{\milli\meter} in size. Water droplets found in raindrops, fogs, hailstones, and snowflakes typically take the shape of oblate spheroids with radii reaching a few tens of millimeters, or nearly perfect spheres with radii below 1 \SI{1}{\milli\meter}. Given that the dimensions of these water droplets align with THz wavelengths (ranging from \SIrange{0.1}{1}{\milli\meter}), they effectively reduce the power of THz waves through absorption and scattering. The ITU-R has provided a power-law equation for modeling rain attenuation based on parameters like distance, rainfall rate (in \si{\milli\meter\per\hour^{}}), and the average size of raindrops \cite{Ref_itu2005specific}. \figurename \ref{Diagram_RainAttenu} shows the rain attenuation described by this ITU-R P838 model from \SIrange{1}{1000}{\giga\hertz} and rain rate from light rain (\SI{1}{\milli\meter\per\hour^{}}) to heavy rain (\SI{200}{\milli\meter\per\hour^{}}).

\subsection{Blockage Loss}
With its diminutive wavelengths in the THz range, surrounding physical objects acquire sufficiently large dimensions for scattering effects, while achieving specular reflections on typical surfaces becomes challenging. Conversely, THz systems heavily rely on narrow pencil beams to enhance the effective propagation distance. Consequently, a LoS path between the transmitting and receiving ends is highly desirable. However, LoS THz links are significantly more prone to obstruction from both macro-scale elements like buildings, furniture, vehicles, and foliage, as well as micro-scale objects including humans, when compared to the conventional sub-6GHz band \cite{Ref_rappaport2014millimeter}. Consequently, it becomes imperative to comprehensively understand the characteristics of blockages and devise effective strategies to prevent blockages or rapidly restore connectivity when a link becomes obstructed. 

\subsection{THz Channel Measurement}
A comprehensive understanding of propagation characteristics and accurate channel models serve as essential prerequisites for formulating transmission algorithms, developing network protocols, and assessing performance in the realm of THz communications and sensing. Channel measurement stands out as the most crucial method for gaining profound insights into THz signal propagation and subsequently establishing precise THz channel models. Given the distinctive attributes of THz signals, cutting-edge sounding equipment becomes imperative for effective channel measurement. Notably, two primary categories of measurement devices are utilized for THz channels: the vector network analyzer (VNA)  and the channel sounder (CS) \cite{Ref_peng2020channel}.

Both these devices acquire channel information by transmitting a reference signal and analyzing the corresponding received signal at the receiver. However, their measurement methodologies differ significantly. The VNA operates within the \textit{frequency domain}, measuring the channel transfer function (CTF) for a specific narrowband channel at each instance and sequentially scanning all frequency points within the designated bandwidth. This approach inherits the benefits of narrowband channel measurements, including high precision due to individual calibration at each frequency point and minimal measurement noise. However, this method is time-consuming and cannot capture dynamic channel variations effectively. On the other hand, the channel-sounding approach functions within the \textit{time domain}, capitalizing on the classical technique known as direct sequence spread spectrum (DSSS). Time-domain channel sounders operate much faster compared to VNAs with frequency scanning, allowing them to capture dynamic changes in the channel. Nevertheless, the precision of time-domain CS is susceptible to significant thermal noise, which scales with the signal bandwidth.

\subsection{THz Channel Modelling}

The development of a wireless communications system necessitates a precise channel model that comprehensively captures the primary propagation attributes corresponding to the operational carrier frequency. This model serves as a valuable tool for wireless researchers and engineers, enabling them to evaluate the effectiveness of diverse transmission algorithms and medium-control protocols, all without resorting to resource-intensive real-world field measurements. In the realm of traditional cellular systems, numerous channel models concentrated on the sub-\SI{6}{\giga\hertz} frequency range have been constructed through techniques such as curve fitting or analytical analysis, drawing from field measurement data. These models effectively encompass all propagation effects, both familiar and novel, resulting in their efficacy. However, given the distinctive nature of THz signal propagation, it becomes imperative to formulate dedicated THz channel models. These models are vital for research, development, performance assessment, and the standardization of THz communications and sensing in 6G. Such tailored channel models are essential due to the unique characteristics of THz signals, facilitating the exploration and advancement of THz technology without solely relying on costly and time-consuming real-world measurements. 

\subsection{THz Antennas}
The traditional principles governing electromagnetic antennas can also find application in the THz range, as pointed out in references \cite{Ref_he2020overview}. Nonetheless, given the considerably high frequencies at play, there emerge specific limitations and disruptive effects demanding careful consideration. The diminutive wavelengths associated with THz frequencies necessitate the creation of exceptionally compact structures, raising concerns about viable manufacturing processes. Conversely, this reduction in structure dimensions opens doors to the adoption of innovative manufacturing methodologies like low-temperature co-fired ceramic (LTCC), antenna-on-chip design, substrate-integrated waveguide (SIW), among others.
Another noteworthy challenge that emerges with escalating frequencies involves the skin effect observed in conductive materials. As the skin depth, denoting the depth of current penetration in a conductive material, diminishes significantly, the conductivity of metallic substances decreases, subsequently amplifying losses within the antenna system.

\subsection{THz Components and Devises}
The THz band was once considered a "terahertz gap" due to its challenging nature. On one hand, \textit{photonic} devices struggle to generate frequencies this low, and on the other, \textit{electronic} oscillators find it arduous to reach such high frequencies. This situation rendered the \si{\THz} band inaccessible for both \textit{photonic} and \textit{electronic} technologies, leading to efforts aimed at generating \si{\THz} radiation from both ends of the spectrum.
Initially, due to component limitations, earlier THz research predominantly focused on applications like imaging or spectrometry \cite{Ref_zandonella2003terahertz}. This was due to two primary factors: These applications demanded relatively high signal output power, but receiver demands were lenient, given that the information carrier was the amplitude rather than the phase of the signal.
System size or specific operational conditions posed no significant restrictions, making certain technologies, especially photonic generation and detection, advantageous.
In contrast, THz communications and sensing require precise phase recovery, especially in cases of digital modulation that leverage both IQ branches. Moreover, compactness and low energy consumption are vital considerations.

\subsection{THz Beamforming}
Leveraging the THz band offers a potential solution to the issue of spectrum scarcity and opens doors to innovative applications like nano-scale networks and in-device communications. However, its practical implementation faces a substantial challenge posed by considerable propagation losses, resulting in notably limited signal transmission distances. This challenge primarily stems from several factors, including the high spreading loss that escalates quadratically with the carrier frequency, gaseous absorption caused by atmospheric oxygen molecules and water vapor, and the detrimental impacts of varying weather conditions, all of which were discussed in the preceding section. Such propagation losses can extend to hundreds of decibels per kilometer or even exceed that threshold. Moreover, this issue is further exacerbated by two additional factors: \cite{Ref_peng2020channel}:
\begin{itemize}
\item \textit{Strong thermal noise}: Noise power is proportional to signal bandwidth with  the constant power density. Therefore, the unique advantage of massive bandwidth at the THz band imposes a side effect of strong thermal noise.  
\item \textit{Hardware constraint}: The transmit power at the THz band is quite constrained since the output power decreases with frequency and is at the level of decibel-milli-Watts in the foreseeable future. Hence,  raising power to extend the communications distance is not feasible.  
\end{itemize}
To extend the signal transmission distance beyond a few meters, high-gain directional antennas and appropriate beamforming approaches are necessary to compensate for such a high propagation loss in THz communications and sensing \cite{Ref_yang2013random, Ref_jiang2012randomBeamforming, Ref_yang2011randomVTC}.

\subsection{THz Beam Alignment}
In response to the formidable propagation losses, the utilization of extensive antenna arrays becomes a viable strategy. Nevertheless, this approach results in the formation of notably focused and narrow beams. To ensure a satisfactory signal-to-noise ratio (SNR) at the receiver and avert disconnections, preserving alignment within degrees is crucial between the transmitter and receiver beams. Hence, achieving beam alignment stands as a pivotal concern essential for establishing a dependable connection. This objective is attained by aligning the beams at both the transmitter and receiver with the direction of the channel paths, wherein channel state information assumes a critical role in facilitating meticulous alignment. However, conventional channel sensing methods employed at lower frequencies become infeasible in the context of THz frequencies. This is attributed to the substantial path losses, rendering pilot signals undetectable during the initial link establishment phase.

\subsection{THz Beam Squint}
Beam squint refers to the phenomenon where the direction of the main lobe or peak of a transmitted or received beam in the 
THz frequency range shifts or changes as a result of altering the frequency or the steering angle of the beamforming system. This shift can be a consequence of various factors, including the properties of the antenna, the design of the beamforming system, and the frequency characteristics. In practical terms, beam squint in the THz range can introduce challenges to maintaining accurate alignment between communicating devices. Given the highly directional nature of THz beams, even slight shifts in the direction of the main lobe can result in misalignment between transmitter and receiver. As a result, precise compensation techniques, dynamic beamforming, and accurate channel state information become crucial to mitigate the effects of THz beam squint and ensure effective communication.

\section{Conclusions}
This paper offered a brief but holistic view of opportunities opened by the under-exploited THz band, including immersive Tbps communications, nano-scale networks, sensing, imaging, and positioning, as well as THz-enabled ISAC. In addition, this paper sheds light on the challenges that must be addressed to materialize the potential of THz-based communications and sensing in 6G systems, such as severe propagation effects, THz channel measurement and modeling, THz hardware, and beam-related issues. We hope this work can highlight the issues and challenges that are open for further research to speed up the research endeavors.

\bibliographystyle{IEEEtran}
\bibliography{IEEEabrv,Ref_ICCC23}

\end{document}